\documentclass[11pt]{article}
\usepackage{amsfonts,amsmath,amssymb}
\usepackage{graphicx}
\usepackage{bm}

\newfont{\twelvemsb}{msbm10 scaled\magstep1}
\newfont{\eightmsb}{msbm8}
\newfam\msbfam
\textfont\msbfam=\twelvemsb \scriptfont\msbfam=\eightmsb
\catcode`\@=11
\def\Bbb{\ifmmode\let\next\Bbb@\else
\def\next{\errmessage{Use \string\Bbb\space only in math mode}}\fi\next}
\def\Bbb@#1{{\fam\msbfam{{#1}}}}

\newcommand{\be}{\begin{equation}}
\newcommand{\ee}{\end{equation}}
\newcommand{\ba}{\begin{eqnarray}}
\newcommand{\ea}{\end{eqnarray}}

\begin{document}
\sloppy
\renewcommand{\thefootnote}{\fnsymbol{footnote}}
\newpage
\setcounter{page}{1} \vspace{0.7cm}

\vspace*{1cm}
\begin{center}
  {\bf On the Lorentz-invariance of the Dyson series \\ in theories
    with derivative couplings}

\vspace{1.8cm} {\large Vincenzo Denisi$^a$, Alessandro Papa$^{a,b}$
and Marco Rossi$^{a,b}$
\footnote{E-mail: vcndns@gmail.com, papa@cs.infn.it, rossi@cs.infn.it}}\\
\vspace{.5cm} $^a${\em Dipartimento di Fisica, Universit\`a della Calabria,\\
I-87036 Arcavacata di Rende, Cosenza, Italy} \\
\vspace{.5cm} $^b${\em Istituto Nazionale di Fisica Nucleare, Gruppo collegato
  di Cosenza,\\  I-87036 Arcavacata di Rende, Cosenza, Italy} \\
\end{center}
\renewcommand{\thefootnote}{\arabic{footnote}}
\setcounter{footnote}{0}
\begin{abstract}
  We study Dyson series for the $S$-matrix when the interaction depends
  on derivatives of the fields. We concentrate on two particular examples: the scalar electrodynamics and the
  renormalised $\phi ^4$ theory. By using Wick's theorem,
  we eventually give evidence that Lorentz invariance is satisfied and that
  usual Feynman rules can be applied to the interaction Lagrangian.

\end{abstract}
\vspace{1cm} 
\newpage

\section{Introduction}

Feynman diagrams and rules are by far the most efficient and convenient
way to build theoretical predictions within field theories amenable of a
perturbative treatment. They manifestly keep Lorentz invariance and come
naturally if field theories are quantised by means of a functional generator
based on the Lagrangian of the theory, which is a Lorentz scalar.
The same holds in the canonical approach, which is based on the Dyson
series for the $S$-matrix in the interaction scheme, for theories not
featuring derivative couplings. In this case, indeed, the interaction
Hamiltonian entering the Dyson series coincides, up to a sign, with the
(scalar) interaction Lagrangian, and Lorentz invariance of the
$S$-matrix, and of the ensuing Feynman rules, is again manifest.

The equivalence between the Feynman approach and the Dyson series is
not evident if the interaction Lagrangian contains derivatives of the fields.
This occurs since the interaction Hamiltonian contains non-invariant terms,
which seem to jeopardize the Lorentz invariance of the $S$-matrix and the
derivation of the usual Feynman rules. This problem was known since the late
40's, when the achievement of a fully covariant formulation of QED
free of divergences at any order in perturbation theory, stimulated the
perturbative investigation and the proof of renormalisability also in
other theories. Among them, {\em scalar} QED received a special attention,
being physically interesting on its own and posing additional technical
problems due to its dependence on derivative couplings. Already in 1950,
Rohrlich~\cite{Rohrlich:1950zz} tackled the problem and showed that, at any
perturbative order, non-invariant terms of the interaction Hamiltonian
are exactly compensated in the Dyson series by non-covariant terms arising
from the time-ordered product
of the derivatives of two fields. The argument by Rohrlich is presented at the
lowest perturbative order in the textbook~\cite{Greiner:1996zu}, whereas
in the textbook~\cite{Itzykson:1980rh} a nice general proof is presented for
the cancellation of non-invariant terms in the Green functions of the theory.

The problems raised by derivative couplings in other theories were
discussed in Refs.~\cite{Matthews,Lee:1962vm}.
A general solution of these problems in the case of quantum mechanics was then proposed by Nambu~\cite{Nambu}.
He proved the equivalence of the Dyson series for the Lagrangian, with a modified $T$-product, in which time derivatives are performed after time ordering, with the Dyson series with standard $T$-products for operators that he eventually proves to be the Hamiltonian. The extension of this proof to QFT was done by Nishijima~\cite{Nishijima}: in this case the modified $T$-product is manifestly covariant and this property eventually allows to prove Lorentz invariance of the Dyson series.
We observe that these results, though conclusive, were obtained without using Wick's theorem.

Other mentionable literature on the same subject are
Refs.~\cite{Sukhanov}, where however the main focus is on
the very definition of interaction Hamiltonian in the interaction scheme,
the issue of non-invariant terms in the $S$-matrix being touched laterally.
A more direct attack to the problem
of derivative couplings by using Wick's theorem can be found in Ref.~\cite{Pavlov}: in brief, supposing
the equivalence of the standard Dyson form for the $S$-matrix,
involving the standard time-ordered product of interaction Hamiltonians,
with the 'Wick form' for the $S$-matrix, involving a modified time-ordered
product of interaction Lagrangians, they find a form for the interaction Hamiltonian.

In this scenario, our aim is to give an alternative proof of the equivalence of the Dyson series involving the Hamiltonian with standard $T$-product
with a Dyson series involving the Lagrangian, with a modified covariant $T$-product. Differently from~\cite{Nishijima}, we derive from first
principles the interaction Hamiltonian and start from usual Dyson series for the Hamiltonian; then, we use Wick's theorem to trade standard $T$-products with normal ordered products and propagators and to eventually express everything in terms of modified covariant $T$-products; after discarding vacuum diagrams, we recognize a Dyson series involving the Lagrangian, with a modified covariant $T$-product.
We believe that the use of Wick's theorem makes plain and
pedagogical the proof of the Lorentz invariance of the Dyson series and its the coincidence with the perturbative series coming from Feynman diagrams. This is the main advantage of our approach. In addition, with our method we can emphasise the fact that such coincidence is true if one discards in the original Dyson series for Hamiltonian vacuum diagrams. This fact was not explicitly remarked in previous literature~\cite{Nambu,Nishijima}.

The plan of the paper is the following. We first
revisit the case of scalar electrodynamics (Section~2) and extend to all orders
the analysis given in~\cite{Greiner:1996zu}, with the aim of
stating the problem in the simplest possible way and to illustrate its
solution. Then we move to the case of the renormalised $\phi^4$ theory
(Section~3), which we take as a representative of all theories where
derivative couplings, not originally present in the bare Lagrangian, appear
due to the renormalisation procedure, as unavoidable counterterms of the
kinetic term. In this theory, now at the lowest orders in perturbation
theory, we address the problem of cancellation of non-invariant terms and
present, as a by-product, a consistent way to define the interaction
Hamiltonian in the interaction scheme.

\section{Dyson series of the $S$-matrix and Feynman rules: the usual
procedure}
\label{setup}
\setcounter{equation}{0}

Let us briefly review how usual perturbative computations based on Feynman
diagrams stem from the Dyson series and Wick theorem.
Here, the word 'usual' refers to the fact that the Lagrangian of the
theory ${\cal L}$, which depends on the fields $\phi _r$ and on their
derivatives $\partial _{\mu} \phi _r$, is decomposed as
\be
   {\cal L}(\phi _r, \partial _{\mu} \phi _r)={\cal L}_0(\phi _r, \partial _{\mu}
   \phi _r) + {\cal L}'(\phi _r) \;,
\label{lagr}
\ee
where ${\cal L}_0$, the free Lagrangian, depends on the fields and their
derivatives, while ${\cal L}'$, the interaction part, depends only on the
fields (not on their derivatives). The free Lagrangian is quadratic in the
fields, whilst the interaction Lagrangian contains terms at least cubic in
the fields and is proportional to a set of real numbers, the coupling constants.
All fields $\phi _r$ in~(\ref{lagr}), and in general all the fields throughout the paper
without any additional index or subscript, are intended to be in Heisenberg representation.

The definition of the momenta,
\be
\pi _r= \frac{\partial {\cal L}}{\partial (\partial _0 \phi _r)}=\frac{\partial
  {\cal L}_0}{\partial (\partial _0 \phi _r)} \, ,
\ee
allows to introduce the Hamiltonian density
\be
{\cal H}(\phi _r, \pi _r)=\pi _r \dot{\phi _r} - {\cal L}= \pi _r \dot{\phi _r} - {\cal L}_0
-{\cal L}'={\cal H}_0+{\cal H}' \, ,
\ee
with
\be
   {\cal H}_0(\phi _r, \pi _r)=\pi _r \dot{\phi _r} - {\cal L}_0 \, , \quad
   {\cal H}'(\phi _r)=-{\cal L}'(\phi _r) \;.
\label{HLint}
\ee
The next ingredient is the Dyson series for the $S$-matrix,
\be
S=\sum_{n=0}^{+\infty} \frac{(-i)^n}{n!}\int d^4x_1\ldots d^4x_n
T \left [ {\cal H}'_I (x_1) \ldots {\cal H}'_I (x_n) \right ] \;,
\label{Dyson}
\ee
which is written in terms of the interaction Hamiltonian in the so-called
interaction representation,
\be
{\cal H}'_I (t)=U{\cal H}' (t) U^{-1} \, , \quad U=e^{iH_0^{(s)}t}e^{-iH^{(s)}t} \, ,
\ee
with $H_0^{(s)}=\int d^{3}x {\cal H}_{0}^{(s)}$ the free Hamiltonian and
$H^{(s)}=\int d^{3}x {\cal H}^{(s)}$ the complete Hamiltonian, both in the
Schr\"odinger representation.
Since ${\cal H}'(\phi _r)=-{\cal L}'(\phi _r)$, one finds that
\be
{\cal H}'_I = {\cal H}'({\phi _r}_I)=-{\cal L}'({\phi _r}_I)
\label{HLintint}\;,
\ee
so that the Dyson series is written as
\be
S=\sum_{n=0}^{+\infty} \frac{i^n}{n!}\int d^4x_1 \ldots d^4x_n
T \left [ {\cal L}'({\phi _r}_I(x_1)) \ldots {\cal L}'({\phi _r}_I(x_n)) \right ] \; ,
\label{SFeynman}
\ee
in terms of the interaction Lagrangian in the Heisenberg scheme, in which all
the fields are in the interaction representation. Through the use of the Wick
theorem, we eventually find that perturbative computations can be organised
by means of the usual Feynman rules applied to $ {\cal L}'$.

Clearly, it seems that this picture collapse when $ {\cal L}'$ contains also
derivatives of the fields $\phi _r$.
What is certainly true in general is that the Dyson series is given
by~(\ref{Dyson}). What is no more true is the second of~(\ref{HLint})
and~(\ref{HLintint}). In addition, the application of the Wick theorem to
the $S$-matrix expansion~(\ref{Dyson}), in which objects inside the $T$-ordered
product depend on derivatives of the fields, is not equivalent to
applying Feynman rules, since
\be
\langle 0 | T[ \partial _\mu \phi (x) \partial _\nu \phi (y)] |0\rangle
  \not= \partial _\mu  \partial _\nu \langle 0 | T[ \phi (x) \phi (y)] |0\rangle
    \;.
\ee
The l.h.s. of this expression is what comes from Wick theorem, the r.h.s. is
what comes from Feynman rules, since in this approach derivatives are attached
to vertices, whilst internal lines are associated to propagators
$\langle 0| T[ \phi (x) \phi (y) |0\rangle $.

However, we will show in two examples, that all these problem 'cancel' each
other: then, using the Wick theorem in the Dyson series~(\ref{Dyson}) is
equivalent to applying Feynman rules to~(\ref{SFeynman}), which contains
$ {\cal L}'({\phi _r}_I)$, {\it i.e.} the interaction Lagrangian with all
the fields in interaction representation.

The two examples we study are scalar electrodynamics and renormalised
$\phi ^4$ theory in four dimensions~\footnote{The arguments presented below
  are in fact independent of the space-time dimension.}.

  \section{Scalar electrodynamics}
\label{scalar_QED}
\setcounter{equation}{0}

As well known, the Lagrangian of scalar electrodynamics is
\begin{equation}
  {\cal L}=\bigl[D_{\mu}\phi\bigr]^\dag D^{\mu} \phi - m^2\phi^\dag\phi
  - \frac{1}{4}F_{\mu\nu}F^{\mu\nu} \;,
\end{equation}
with
\begin{equation}
D_{\mu}=\partial_{\mu} - ieA_{\mu} \;.
\end{equation}
Then, we have ${\cal L}={\cal L}_0+{\cal L}'$, with
\begin{equation}
  {\cal L}'=ieA_{\mu}\phi^\dag\partial^{\mu}\phi-ieA^{\mu}
  \bigl(\partial_{\mu}\phi^\dag\bigr)\phi+e^2A_{\mu}A^{\mu}\phi^\dag\phi \;.
\label{Linter}
\end{equation}
Defining the conjugate fields
\begin{equation}
  \pi=\frac{\partial {\cal L}}{\partial\dot{\phi}}=\dot{\phi}^\dag+ieA_{0}
  \phi^\dag \;, \qquad  \pi^\dag
  =\frac{\partial {\cal L}}{\partial\dot{\phi}^\dag}=\dot{\phi}-ieA^{0}
  \phi \;,
\end{equation}
we introduce the Hamiltonian density
\begin{equation}
{\cal H}=\pi\dot{\phi}+\dot{\phi}^\dag\pi^\dag-{\cal L} \;,
\end{equation}
which we write ${\cal H}={\cal H}_0+{\cal H}'$, where
\be
   {\cal H}_{0}=\pi^\dag\pi +{\bm{\nabla}}\phi^\dag {\bm{\nabla}}\phi
   + m^2\phi^\dag\phi  + \frac{1}{4}F_{\mu\nu}F^{\mu\nu}
\ee
and
\be
   {\cal H}'=-ieA_{0}\phi^\dag(\pi^\dag+ieA_0\phi)-ie\textbf{A}\phi^\dag
   {\bm{\nabla}}\phi
   +ieA_{0}(\pi-ieA_0\phi^\dag)\phi+ie\textbf{A}({\bm{\nabla}}\phi^\dag)\phi
   -e^2A_{\mu}A^{\mu}\phi^\dag\phi -e^2A_{0}^{2}\phi^\dag\phi \;.
\ee
So far, all the expressions above are in Heisenberg representation. Operators
with no subscript are in Heisenberg representation. Moreover, it is understood
that all terms in the Lagrangian and Hamiltonian densities are subject
to normal ordering, $N$. In order to write the Dyson
series we have to pass to the interaction representation. We find useful the
following property
\begin{equation}
\begin{split}
&U\pi^{\dag}(x)U^{-1}=\partial^0\phi_I(x) \;,\\
\label{heis-int}\\
&U\pi (x)U^{-1}=\partial^0\phi^{\dag}_I(x) \;,
\end{split}
\end{equation}
where the operator
\be
U=e^{iH_0^{(s)}t}e^{-iH^{(s)}t} \;,
\ee
with $H_0^{(s)}=\int d^{3}x {\cal H}_{0}^{(s)}$ the free Hamiltonian and
$H^{(s)}=\int d^{3}x {\cal H}^{(s)}$ the complete Hamiltonian, both in the
Schr\"odinger representation, allows to pass from Heisenberg to interaction
representation. We give a proof of~(\ref{heis-int}) in Appendix~\ref{appA}.
Using~(\ref{heis-int}), we obtain for the interaction Hamiltonian in the
interaction representation the following expression:
\be
U{\cal H}'U^{-1}={\cal H}'_{I}=-ie{A_{\mu}}_I\phi^\dag_I\partial^{\mu}\phi_I+ieA^{\mu}_I
\bigl(\partial_{\mu}\phi^\dag_I\bigr)\phi_I-e^2{A_{\mu}}_IA^{\mu}_I\phi^\dag_I\phi_I
+e^2{A_{0}}^{2}_I\phi^\dag_I\phi_I \;,
\label{Hintint}
\ee
To simplify notations, in the following we remove the subscript $_I$ in all the fields, because from now
on all the fields are in the interaction representation. However, we keep the
subscript $_I$ in the Hamiltonian, to stress that it is in the interaction
representation.
Comparing~(\ref{Hintint}) with the interaction Lagrangian~(\ref{Linter}), in
which all the fields are promoted to be in interaction representation~\footnote{Actually, this procedure is in our opinion the correct one to define (interaction) Lagrangians in interaction representation in a general case.}, we find
that
\be
   {\cal H}'_{I}=-{\cal L}'+e^2A_{0}^{2}\phi^\dag\phi
   \equiv -{\cal L}'+{\cal R}  \;.
\label{HLrel}
\ee
We did not put the index $_I$ in the Lagrangian in~(\ref{HLrel}), because it
still has the form of the interaction Lagrangian in Heisenberg representation:
the only caveat, as written before, is that the fields appearing in its
expression~(\ref{Linter}) are in interaction representation.

\medskip

Now we prove the following equality. The Dyson series
\be
S=\sum_{n=0}^{+\infty} \frac{(-i)^n}{n!}\int d^4x_1 \ldots d^4 x_n
T \left [ {\cal H}'_I (x_1) \ldots {\cal H}'_I (x_n) \right ]
\label{Dyson2}
\ee
can be written as
\be
S=\sum_{n=0}^{+\infty} \frac{i^n}{n!}\int d^4x_1 \ldots d^4x_n
\hat T \left [ {\cal L}'(x_1)\ldots {\cal L}'(x_n) \right ] \, ,
\label{SFeynman2}
\ee
provided that we use in~(\ref{SFeynman2}) a modified definition of the
$T$-product. Given
\begin{equation}
  \langle 0 |T\bigl( \phi_1\phi_{2}^{\dag}\bigr)|0\rangle \equiv
  i\Delta_{F}(x_1-x_2) \;,
\end{equation}
the operation $\hat T$ satisfies the Wick theorem, but its 'action' on
elementary fields is the following~\footnote{The $\hat T$-product is known
  in the literature as 'Wick $T$-product', whereas the standard $T$-product
  is called also 'Dyson $T$-product'.}:
\ba
&& \langle 0 | \hat T\bigl( \phi_1\phi_{2}^{\dag}\bigr)|0\rangle
= \langle 0 | T\bigl( \phi_1\phi_{2}^{\dag}\bigr)|0\rangle
\equiv i\Delta_{F}(x_1-x_2) \, , \label{hatT1}\\
&& \langle 0 | \hat T\bigl( (\partial ^{\mu}\phi_1) \phi_{2}^{\dag}\bigr)|0\rangle
\equiv i \partial ^{\mu}_1 \Delta_{F}(x_1-x_2) \, , \label{hatT2} \\
&& \langle 0 | \hat T\bigl( (\partial ^{\mu}\phi_1)(\partial ^{\nu}\phi_{2}^{\dag})
\bigr)|0\rangle \equiv i \partial ^{\mu}_1 \partial ^{\nu}_2 \Delta_{F}(x_1-x_2)
\;. \label{hatT3}
\ea
For the sake of brevity, we have introduced here the notation $\phi_i\equiv
\phi(x_i)$ and $\phi_i^\dagger \equiv \phi^\dagger(x_i)$, as well as
$\partial ^{\mu}_i \equiv \partial/\partial x_{i,\mu}$; below, we will use
similarly $A_{\mu,i}$ for $A_\mu(x_i)$ and will extend this notation
also to functions of fields, as the Lagrangian density.
We remark that the use of the Wick theorem in the expansion~(\ref {SFeynman2})
with the operation $\hat T$ produces a Dyson series whose terms are all
manifestly Lorentz invariant.
Lorentz invariance is not evident using the expansion~(\ref {Dyson2}), which
however is the {\it a priori} correct one.

We now give a perturbative proof of this statement. Let us write the first two
terms of the Dyson series:
\begin{equation}
\begin{split}
S^{(1)}&=-i \int {d^4x_1\Bigl[ {\cal H}'(x_1)\Bigr]}\\
&=-i \int {d^4x_1\Bigl( -{\cal L}' + {\cal R} \Bigr)_1} \;,
\end{split}
\end{equation}
\begin{equation}
\begin{split}
  S^{(2)}&=-\frac{1}{2} \int {d^4x_1 d^4x_2 T\Bigl[ {\cal H}'(x_1){\cal H}'(x_2)
      \Bigr]}\\
  &=-\frac{1}{2} \int {d^4x_1 d^4x_2 T\Bigl[\Bigl(-{\cal L}'+ {\cal R}\Bigr)_{1}
      \Bigl(-{\cal L}' + {\cal R}\Bigr)_{2}\Bigr]} \\
  &=-\frac{1}{2} \int {d^4x_1 d^4x_2 T\Bigl[\Bigl( -{\cal L}' \Bigr)_1
      \Bigl( -{\cal L}'\Bigr)_2 }\\
    &+ {\cal R}_{1} \Bigl( -{\cal L}'\Bigr)_2
    +\Bigl( -{\cal L}' \Bigr)_1 {\cal R}_{2} + {\cal R}_{1} {\cal R}_{2}\Bigr]
    \;.
      \label{S2}
\end{split}
\end{equation}
We remark that in $S^{(1)}$ there is an extra term with respect to $-{\cal L}'$:
$e^2\bigl(A_{0,1}^{2}\bigr)\phi_{1}^\dag\phi_1$, which is not Lorentz invariant.
However, this is not the end of the story, since another source of Lorentz
non-invariance comes from the operation of $T$ arising in various terms of
$S^{(2)}$ after application of the Wick theorem. To be precise we have
that, remembering (\ref{hatT1})-(\ref{hatT3}),
\ba
 \langle 0 | T\bigl( (\partial ^{\mu}\phi_1) \phi_{2}^{\dag}\bigr)|0\rangle
&=&  i \partial ^{\mu}_1 \Delta_{F}(x_1-x_2)
 =\langle 0 | \hat T\bigl( (\partial ^{\mu}\phi_1 ) \phi_{2}^{\dag}\bigr)|0\rangle
 \label{That1} \\
\langle 0 |T\bigl( (\partial ^{\mu}\phi_1)(\partial ^{\nu} \phi_{2}^{\dag})\bigr)
|0\rangle
&=&  i \partial ^{\mu}_1 \partial ^{\nu}_2 \Delta_{F}(x_1-x_2)
-i\delta^{\mu}_{0}\delta^{\nu}_{0}\delta^{(4)}{(x_1-x_2)} \nonumber \\
&=& \langle 0 | \hat T\bigl( (\partial ^{\mu}\phi_1)(\partial ^{\nu}
\phi_{2}^{\dag})\bigr)|0\rangle
-i\delta^{\mu}_{0}\delta^{\nu}_{0}\delta^{(4)}{(x_1-x_2)} \;.
\label{That2}
\ea
We see that a non-covariant term appears in the 'contraction' between
$\partial ^{\mu}\phi_1$ and $\partial ^{\nu} \phi_{2}^{\dag}$. The use of
(\ref{That1}) and~(\ref{That2}) in the term
$T\left[\left( -{\cal L}' \right)_1 \left( -{\cal L}'\right)_2 \right]$ in
$S^{(2)}$ eventually gives
\be
 -\frac{1}{2} \int {d^4x_1 d^4x_2 T\Bigl[\Bigl( -{\cal L}' \Bigr)_1
    \Bigl( -{\cal L}'\Bigr)_2\Bigr]}=
-\frac{1}{2} \int {d^4x_1 d^4x_2 \hat T\Bigl[\Bigl( -{\cal L}' \Bigr)_1
    \Bigl( -{\cal L}'\Bigr)_2 \Bigr]}
+ i\int d^4x_1  {\cal R}_1 \;. \label{wick2}
\ee
Indeed,
\ba
T\Bigl[\Bigl( -{\cal L}' \Bigr)_1 \Bigl( -{\cal L}'\Bigr)_2\Bigr]
&=& T\Bigl[ \Bigl( ieA_{\mu}\phi^\dag\partial^{\mu}\phi-ieA^{\mu}
  \bigl(\partial_{\mu}\phi^\dag\bigr)\phi \Bigr)_1
  \Bigl( ieA_{\nu}\phi^\dag\partial^{\nu}\phi-ieA^{\nu}
  \bigl(\partial_{\nu}\phi^\dag\bigr)\phi \Bigr)_2  \Bigr]+O(e^3) \nonumber \\
&=& T\Bigl[ \Bigl( ieA_{\mu}\phi^\dag\partial^{\mu}\phi \Bigr)_1
  \Bigl( ieA_{\nu}\phi^\dag\partial^{\nu}\phi \Bigr)_2  \Bigr]\nonumber \\
&+& T\Bigl[ \Bigl( ieA_{\mu}\phi^\dag\partial^{\mu}\phi \Bigr)_1
  \Bigl( -ieA^{\nu} \bigl(\partial_{\nu}\phi^\dag\bigr)\phi \Bigr)_2  \Bigr]
\nonumber \\
&+& T\Bigl[ \Bigl( -ieA^{\mu} \bigl(\partial_{\mu}\phi^\dag\bigr)\phi \Bigr)_1
  \Bigl( ieA_{\nu}\phi^\dag\partial^{\nu}\phi \Bigr)_2  \Bigr] \nonumber \\
&+& T\Bigl[ \Bigl( -ieA^{\mu} \bigl(\partial_{\mu}\phi^\dag\bigr)\phi \Bigr)_1
  \Bigl( -ieA^{\nu} \bigl(\partial_{\nu}\phi^\dag\bigr)\phi \Bigr)_2  \Bigr]
+O(e^3) \nonumber \, .
\ea
Now, we use Wick theorem and write the resulting expression in terms of the modified $T$-product,
$\hat T$, by means of (\ref{That1}),~(\ref{That2}). We get
\ba
T\Bigl[\Bigl( -{\cal L}' \Bigr)_1 \Bigl( -{\cal L}'\Bigr)_2\Bigr]
&=& \hat T\Bigl[ \Bigl( ieA_{\mu}\phi^\dag\partial^{\mu}\phi \Bigr)_1
  \Bigl( ieA_{\nu}\phi^\dag\partial^{\nu}\phi \Bigr)_2  \Bigr]\nonumber \\
&+& \hat T\Bigl[ \Bigl( ieA_{\mu}\phi^\dag\partial^{\mu}\phi \Bigr)_1
  \Bigl( -ieA^{\nu} \bigl(\partial_{\nu}\phi^\dag\bigr)\phi \Bigr)_2  \Bigr]
-i e^2 N\bigl[\bigl(A_{0,1}^2\bigr)\phi_{1}^\dag \phi_1 \bigr] \delta^{(4)}(x_1-x_2)
\nonumber \\
&+& \hat T\Bigl[ \Bigl( -ieA^{\mu} \bigl(\partial_{\mu}\phi^\dag\bigr)\phi \Bigr)_1
  \Bigl( ieA_{\nu}\phi^\dag\partial^{\nu}\phi \Bigr)_2  \Bigr]
-i e^2 N\bigl[\bigl(A_{0,1}^2\bigr)\phi_{1}^\dag \phi_1 \bigr] \delta^{(4)}(x_1-x_2)
\nonumber \\
&+& \hat T\Bigl[ \Bigl( -ieA^{\mu} \bigl(\partial_{\mu}\phi^\dag\bigr)\phi \Bigr)_1
  \Bigl( -ieA^{\nu} \bigl(\partial_{\nu}\phi^\dag\bigr)\phi \Bigr)_2  \Bigr]
+O(e^3) \;, \nonumber \\
&=&  \hat T\Bigl[\Bigl( -{\cal L}' \Bigr)_1 \Bigl( -{\cal L}'\Bigr)_2\Bigr]
-2i e^2 N\bigl[\bigl(A_{0,1}^2\bigr)\phi_{1}^\dag \phi_1 \bigr]
\delta^{(4)}(x_1-x_2) \nonumber \\
&=&  \hat T\Bigl[\Bigl( -{\cal L}' \Bigr)_1 \Bigl( -{\cal L}'\Bigr)_2\Bigr]
-2i {\cal R}_1 \delta^{(4)}(x_1-x_2) \;. \label{LL}
\ea

Using~(\ref{LL}) we conclude that
\ba
S^{(1)}+S^{(2)}&=&-i \int {d^4x_1\Bigl( -{\cal L}' \Bigr)_1}-\frac{1}{2}
\int d^4x_1 d^4x_2 \hat T\Bigl[\Bigl( -{\cal L}' \Bigr)_1
    \Bigl( -{\cal L}'\Bigr)_2  \Bigr. \nonumber \\
    &+& \Bigl. {\cal R}_{1} \Bigl( -{\cal L}'\Bigr)_2
    +\Bigl( -{\cal L}' \Bigr)_1 {\cal R}_{2} + {\cal R}_{1} {\cal R}_{2}\Bigr]
    \;, \label{S1+S2}
\ea
where we have used that
\[
T \Bigl({\cal R}_{1} \Bigl( -{\cal L}'\Bigr)_2\Bigr) =
\hat T \Bigl({\cal R}_{1} \Bigl( -{\cal L}'\Bigr)_2\Bigr) \;,
\]
\[
T \Bigl( \Bigl( -{\cal L}' \Bigr)_1 {\cal R}_{2} \Bigr) =
\hat T \Bigl( \Bigl( -{\cal L}' \Bigr)_1 {\cal R}_{2} \Bigr) \;,
\]
\[
T\Bigl( {\cal R}_{1} {\cal R}_{2}\Bigr) =
\hat T\Bigl( {\cal R}_{1} {\cal R}_{2}\Bigr)\;,
\]
since ${\cal R}$ does not contain terms with time derivatives. In other
words, as follows from (\ref {That1}),~(\ref {That2}), the $T$-product differs from the $\hat T$-product only when it
applies to two interaction Lagrangians. This implies that, when applying the
Wick theorem to higher order terms of the Dyson $S$-matrix expansion, the only
source of Lorentz non-invariant terms will be the contraction of two
interaction Lagrangians.

We observe that in~(\ref{S1+S2}) the non-invariant term ${\cal R}_1$,
originally present in $S^{(1)}$, has been canceled by non-invariant term
generated in $S^{(2)}$ by the contraction of the two interaction Lagrangians --
see~(\ref{wick2}).

In the sum $S^{(1)}+S^{(2)}$ there are, however, two terms left which are
non-invariant:
\be
-\frac{1}{2} \int d^4x_1 d^4x_2 \hat T \Bigl[ {\cal R}_{1} \Bigl( -{\cal L}'
  \Bigr)_2 +\Bigl( -{\cal L}' \Bigr)_1 {\cal R}_{2} \Bigr]
= - \int d^4x_1 d^4x_2 \hat T \Bigl[ {\cal R}_{1} \Bigl( -{\cal L}'
  \Bigr)_2 \Bigr] \label{NI1}
\ee
and
\be
-\frac{1}{2} \int d^4x_1 d^4x_2 \hat T\Bigl[{\cal R}_{1} {\cal R}_{2}\Bigr] \;.
\label{NI2}
\ee
The first of these terms is canceled by the non-invariant contributions which
arise in
\[
S^{(3)}=\frac{(-i)^3}{3!} \int {d^4x_1 d^4x_2 d^4x_3
  T\Bigl[\Bigl(-{\cal L}'+{\cal R} \Bigr)_1
         \Bigl(-{\cal L}'+{\cal R} \Bigr)_2
         \Bigl(-{\cal L}'+{\cal R} \Bigr)_3 \Bigr]}
\]
from terms with three ${\cal L}'$s, two of them being contracted. There
are three equivalent such terms, which, recalling~({\ref{LL}), are easily
shown to sum up to
\[
\frac{(-i)^3}{3!} \int d^4x_1 d^4x_2 \ 3 \ \hat T\Bigl[-2i {\cal R}_1
    \Bigl(-{\cal L}'\Bigr)_2\Bigr]
\]
and, therefore, exactly cancel~(\ref{NI1}) in $S^{(1)}+S^{(2)}+S^{(3)}$.
The other non-invariant term of $S^{(1)}+S^{(2)}$, given in~(\ref{NI2}),
is cancelled in $S^{(1)}+S^{(2)}+S^{(3)}+S^{(4)}$ by the three equivalent terms
in $S^{(3)}$ with one ${\cal R}$ and two contracted ${\cal L}'$s
and by the three equivalent terms in $S^{(4)}$ with four ${\cal L}'$s
pairwise contracted.

This pattern of cancellations can be generalized. Non-invariant terms
containing $n$ factors of the type ${\cal R}$ and $m$ factors of the
type ${\cal L}'$, which can always be put in the form
\be
\hat T \Bigl[{\cal R}_1 \ldots {\cal R}_n \Bigl(-{\cal L}'\Bigr)_{n+1}
  \ldots \Bigl(-{\cal L}'\Bigr)_{n+m} \Bigr] \label{NIgen}\;,
\ee
appear first in $S^{(n+m)}$ and arise also in $S^{(n+m+1)}$ (in terms with
$n-1$ factors of the type ${\cal R}$ and {\em one} pair of contracted
${\cal L}'$s), then in $S^{(n+m+2)}$ (in terms with $n-2$ factors of the type
${\cal R}$ and {\em two} pairs of contracted ${\cal L}'$s), {\it etc}. The last
appearance is in $S^{(2n+m)}$, in terms with no factors of the type ${\cal R}$
and $n$ pairs of contracted ${\cal L}'$s. To summarize, ({\ref{NIgen})
  appears in $S^{(n+m+j)}$, $j=0,1, \ldots, n$, in terms with $n-j$ factors ${\cal R}$
  and $j$ contractions of interaction Lagrangian pairs. Each contraction brings along
  a Dirac delta which cancels one of the integrations over the space-time, so
  that all terms end up to integrated as in $S^{(n+m)}$, {\it i.e.} over
  $d^4x_1 \ldots d^4x_{n+m}$. The combinatorial weight in
  which the term~({\ref{NIgen}) appears in $S^{(n+m+j)}$ is given by
\[
w^{nm}_j \equiv \frac{(-i)^{n+m+j}}{(n+m+j)!} \ \binom{n+m+j}{m+2j}
\ \binom{m+2j}{2j} \ (2j-1)!! \ (-2i)^j \;,
\]
where the first factor comes from the definition of Dyson series, the
second counts the number of (equivalent) terms in $S^{(n+m+j)}$ with
$n-j$ factors of type ${\cal R}$ and $m+2j$ factors of type ${\cal L}'$,
the third counts the number of ways to select $2j$ Lagrangians to be contracted
out of the $m+2j$ available ones, the fourth is the number of ways $2j$
Lagrangians can be pairwise contracted, the last factor comes from the fact
that each of the $j$ contractions of two Lagrangians gives $-2i {\cal R}$.
The total weight of the non-invariant term~(\ref{NIgen}) is therefore
\[
\sum_{j=0}^n w^{nm}_j
= \frac{(-i)^{n+m}}{m!} \ \sum_{j=0}^n (-2)^j \frac{(2j-1)!!}{(2j)!(n-j)!} =0 \;,
\]
since, observing that $(2j)!= 2^j j! (2j-1)!!$, we have
\[
\sum_{j=0}^n (-2)^j \frac{(2j-1)!!}{(2j)!(n-j)!}
= \frac{1}{n!} \sum_{j=0}^n (-1)^j \binom{n}{j} = \frac{1}{n!} (1-1)^n=0\;.
\]
We have then proved that Dyson series (\ref {Dyson2}) can be traded for the manifestly Lorentz invariant series
(\ref{SFeynman2}). Perturbative expansion for (\ref{SFeynman2}) can then be organised according to usual Feynman rules for scalar electrodynamics.

\section{Renormalised $\phi ^4$ theory}
\label{phi4}
\setcounter{equation}{0}

We consider here the theory of a massless real scalar field, undergoing a
quartic self-interaction, as a simple representative of all field theories which
acquire an interaction term depending on derivatives of the fields through
the procedure of perturbative renormalisation~\footnote{The actual perturbative
  renormalisability of the $\phi^4$ theory and the triviality issue are
  inessential in this context.}.

We will show that a mechanism of cancellation of non-covariant terms takes
place on similar grounds as for scalar electrodynamics, modulo a couple
of caveats which make the present case interesting {\it per s\'e}.

The Lagrangian of the theory is
\begin{equation}
  {\cal L}=\frac{1}{2}\partial_\mu \phi \partial^\mu \phi - \frac{\lambda}{4!}
  \phi^4 \;.
  \label{Lbare}
\end{equation}
The starting step of perturbative renormalisation is to redefine field and
coupling as
\[
\phi = Z^{1/2}\phi_{\rm R}\;,
\]
\[
\lambda = Z_\lambda \lambda_{\rm R}\;,
\]
leading to the following expression for the Lagrangian:
\begin{equation}
  {\cal L}=\frac{Z}{2}\partial_\mu \phi_{\rm R} \partial^\mu \phi_{\rm R}
  - \frac{Z^2 Z_\lambda \lambda_{\rm R}}{4!} \phi_{\rm R}^4 \;,
\end{equation}
which can be recast in the form
\ba
  {\cal L}&=&\frac{1}{2}\partial_\mu \phi_{\rm R} \partial^\mu \phi_{\rm R}
    - \frac{\lambda_{\rm R}}{4!} \phi_{\rm R}^4 \nonumber \\
&+& \frac{Z-1}{2}\partial_\mu \phi_{\rm R} \partial^\mu \phi_{\rm R}
  - \frac{(Z^2 Z_\lambda -1) \lambda_{\rm R}}{4!} \phi_{\rm R}^4 \;.
\label{Lren}
\ea
The first two terms in ${\cal L}$ have the same form as in the original Lagrangian,
but they are written through renormalised field and coupling; the remaining two
terms are the so-called 'counterterms'. For the purposes of perturbative
calculations, and of the related renormalisation procedure, all terms but the
first one in~(\ref{Lren}) must be considered as interaction terms, so that
we can write ${\cal L}={\cal L}_0+{\cal L}'$, with
\be
  {\cal L}_0 = \frac{1}{2}\partial_\mu \phi \partial^\mu \phi
  \label{L0}
\ee
and
\be
  {\cal L}' = - \frac{\lambda}{4!} \phi^4
+ \frac{Z-1}{2}\partial_\mu \phi \partial^\mu \phi
  - \frac{(Z^2 Z_\lambda -1) \lambda}{4!} \phi^4 \;,
   \label{Lint}
\ee
where we have omitted for brevity the subscript $R$, understanding that, from now
on, field and coupling are always the renormalised ones. We can see that
${\cal L}'$ contains an interaction term depending on the field derivatives
in spite of the fact that the original 'bare' theory had a derivative-free
interaction. Moreover, ${\cal L}'$ depends on the renormalised coupling
$\lambda$ both explicitly and through the renormalisation constants
$Z$ and $Z_\lambda$, which in perturbation theory must take the form of
a power series in $\lambda$, the constant term being equal to one. In the
following, it will prove convenient to consider $(Z-1)$ and
$(Z^2 Z_\lambda -1)$ as additional, independent couplings, their relation
to $\lambda$, i.e. the fact that they are both $O(\lambda)$,
being used only to justify their smallness and, therefore,
their suitability as expansion parameters.

To introduce the Hamiltonian, we have to define the conjugate field:
\begin{equation}
  \pi=\frac{\partial {\cal L}}{\partial\dot{\phi}}=Z \dot{\phi}\;.
\end{equation}
We stress that $\phi$ here is the renormalised field, therefore $\pi$,
after quantisation, will implicitly enter the canonical commutation relations
together with $\phi$. It can be easily shown that the equations of motion
for $\phi$ and $\pi$, as derived from their commutators with the Hamiltonian
(to be written below), are equivalent to the equation of motion for the bare
field, {\it i.e.} the Euler-Lagrange equation for the bare field $\phi$ as
derived from the original Lagrangian~(\ref{Lbare}). This is in marked
contrast with Ref.~\cite{ING}, where instead canonical commutation relations
were imposed at the level of the bare fields and an {\it ad hoc} modification
of the Hamiltonian had to be performed to obtain the equation of motion
of the bare field from the Hamiltonian dynamics.

The Hamiltonian density is defined in the usual way:
\ba
   {\cal H}&=&\pi\dot{\phi}-{\cal L} \\
   &=& \frac{\pi^2}{2}+\frac{1}{2}(\bm{\nabla}\phi)^2
   +\frac{\lambda}{4!}\phi^4 -\frac{\pi^2(Z-1)}{2Z}
   +\frac{(Z-1)}{2} (\bm{\nabla}\phi)^2
   +\frac{(Z^2Z_{\lambda}-1)\lambda}{4!}\phi^4 \;,
\ea
which we can split as ${\cal H}={\cal H}_0+{\cal H}'$, with
\be
   {\cal H}_{0}=\frac{\pi^2}{2}+\frac{1}{2}(\bm{\nabla}\phi)^2
\ee
and
\be
   {\cal H}' = \frac{\lambda}{4!}\phi^4 -\frac{\pi^2(Z-1)}{2Z}
   +\frac{(Z-1)}{2} (\bm{\nabla}\phi)^2 +\frac{(Z^2Z_{\lambda}-1)\lambda}{4!}
   \phi^4 \;.
\ee
So far, all the fields are in Heisenberg representation and, again,
all terms in the Lagrangian and Hamiltonian densities are implicitly
assumed to be subject to normal ordering, $N$. In order to write the Dyson
series we have to pass to the interaction representation and can use the
property
\begin{equation}
  U\pi(x)U^{-1}=\partial^0\phi_I(x) \;,
  \label{heis-int-scal}
\end{equation}
which is analogous to~(\ref{heis-int}) for a real scalar field.
Using~(\ref{heis-int-scal}), we obtain for the interaction Hamiltonian
in the
interaction representation the following expression:
\be
U{\cal H}'U^{-1}={\cal H}'_{I}=\frac{\lambda}{4!}\phi^4
-\frac{\dot{\phi}^2(Z-1)}{2Z} +\frac{(Z-1)}{2} (\bm{\nabla}\phi)^2
+\frac{(Z^2Z_{\lambda}-1)\lambda}{4!}\phi^4\; ,
\label{Hintint_phi}
\ee
where all the fields are to be intended in interaction representation. In
the following we remove the subscript $_I$ in all the fields, because from now
on all the fields are in the interaction representation. However, we keep the
subscript $_I$ in the Hamiltonian, to stress that it is in the interaction
representation. Comparing~(\ref{Hintint_phi}) with the interaction
Lagrangian~(\ref{Lint}), in which all the fields are promoted to be in interaction representation,
we find
that
\be
   {\cal H}'_{I}=-{\cal L}'+\frac{(Z-1)^2}{2Z}\dot{\phi}^2 \;.
   \label{HLrel_phi}
\ee
We did not put the subscript $_I$ in the Lagrangian in~(\ref{HLrel_phi}), because
it still has the form of the interaction Lagrangian in Heisenberg
representation: the only caveat, as written before, is that the fields
appearing in its expression~(\ref{Lint}) are in interaction
representation. We observe that ${\cal H}'_{I}$ is not Lorentz-invariant,
due to the presence of the term depending on $\dot{\phi}$. This expression
for ${\cal H}'_{I}$ agrees, {\it mutatis mutandis}, with the one found
in Ref.~\cite{ING}.

The stage now is set to prove that, also in the present case, the Dyson
series~(\ref{Dyson2}) can be written as in~(\ref{SFeynman2}), provided
that a modified definition of the $T$-product, $\hat T$, is used: as before, $\hat T$ satisfies Wick theorem and
\ba
&& \langle 0 | \hat T\bigl( \phi_1\phi_{2}\bigr)|0\rangle
= \langle 0 | T\bigl( \phi_1\phi_{2}\bigr)|0\rangle
\equiv i\Delta_{F}(x_1-x_2) \\
&& \langle 0 | \hat T\bigl( (\partial ^{\mu}\phi_1) \phi_{2}\bigr)|0\rangle
\equiv i \partial ^{\mu}_1 \Delta_{F}(x_1-x_2) \\
&& \langle 0 | \hat T\bigl( (\partial ^{\mu}\phi_1)(\partial ^{\nu}\phi_{2})
\bigr)|0\rangle \equiv i \partial ^{\mu}_1 \partial ^{\nu}_2 \Delta_{F}(x_1-x_2)
\;.
\ea

We present a sketch of the perturbative proof of the validity of the
expansion~(\ref{SFeynman2}). The two main ingredients are, as in the case
studied in the previous Section, the Wick theorem and the following relations
between the standard $T$-product and the modified one, $\hat T$:
\ba
 \langle 0 | T\bigl( (\partial ^{\mu}\phi_1) \phi_{2}\bigr)|0\rangle
&=&  i \partial ^{\mu}_1 \Delta_{F}(x_1-x_2)
 =\langle 0 | \hat T\bigl( (\partial ^{\mu}\phi_1 ) \phi_{2}\bigr)|0\rangle
 \label{That1_phi} \\
\langle 0 |T\bigl( (\partial ^{\mu}\phi_1)(\partial ^{\nu} \phi_{2})\bigr)
|0\rangle
&=&  i \partial ^{\mu}_1 \partial ^{\nu}_2 \Delta_{F}(x_1-x_2)
-i\delta^{\mu}_{0}\delta^{\nu}_{0} \delta^{(4)}{(x_1-x_2)} \nonumber \\
&=& \langle 0 | \hat T\bigl( (\partial ^{\mu}\phi_1)(\partial ^{\nu}
\phi_{2})\bigr)|0\rangle
-i\delta^{\mu}_{0}\delta^{\nu}_{0} \delta^{(4)}{(x_1-x_2)} \;.
\label{That2_phi}
\ea

Let us then write the first term of the Dyson series:
\ba
S^{(1)}&=&-i \int {d^4x_1\Bigl[ {\cal H}'(x_1)\Bigr]} \\
&=&-i \int {d^4x_1\Bigl( -{\cal L}' + \frac{(Z-1)^2}{2Z}\dot{\phi}^2 \Bigr)_1}
\;. \nonumber
\ea
Consider that
\[
\frac{(Z-1)^2}{2Z}= \frac{(Z-1)^2}{2} \frac{1}{1+(Z-1)}
=\frac{(Z-1)^2}{2} \Bigl[1-(Z-1)+(Z-1)^2-(Z-1)^3+\ldots\Bigr]\;,
\]
where each term is proportional to an integer power of the 'coupling' $(Z-1)$,
starting from $(Z-1)^2$. This means that, to cancel all non-invariant terms in
$S^{(1)}$, one needs to consider the non-invariant terms arising from the
operation of $T$ through the Wick theorem in {\em all} other pieces $S^{(n)}$
of the Dyson expansion. Let us work this out explicitly for the lowest-order
contribution, proportional to $(Z-1)^2$, which requires considering, in
addition to $S^{(1)}$, just $S^{(2)}$:
\begin{equation}
\begin{split}
  S^{(2)}&=-\frac{1}{2} \int {d^4x_1 d^4x_2 T\Bigl[ {\cal H}'(x_1){\cal H}'(x_2)
      \Bigr]}\\
  &=-\frac{1}{2} \int {d^4x_1 d^4x_2 T\Bigl[\Bigl( -{\cal L}'
      + \frac{(Z-1)^2}{2Z}\dot{\phi}^2\Bigr)_{1} \Bigl( -{\cal L}'
      + \frac{(Z-1)^2}{2Z}\dot{\phi}^2\Bigr)_{2}\Bigr]} \;.
\end{split}
\end{equation}
Restricting to contributions at most of order $(Z-1)^2$, we notice that
\ba
T\Bigl[\Bigl( -{\cal L}' \Bigr)_1 \Bigl( -{\cal L}'\Bigr)_2\Bigr]
&=& T\Bigl[ \Bigl( -\frac{\lambda}{4!}\phi^4 + \frac{(Z-1)}{2}
\partial_{\mu}\phi\partial^{\mu}\phi-\frac{(Z^2Z_{\lambda}-1)\lambda}{4!}\phi^4
\Bigr)_1 \Bigr. \nonumber \\
&\times &\Bigl( -\frac{\lambda}{4!}\phi^4 + \frac{(Z-1)}{2}
\partial_{\nu}\phi\partial^{\nu}\phi-\frac{(Z^2Z_{\lambda}-1)\lambda}{4!}\phi^4
\Bigr)_2 \Bigr] \nonumber \\
&=& T\Bigl[ \Bigl( -\frac{\lambda}{4!}\phi^4
  -\frac{(Z^2Z_{\lambda}-1)\lambda}{4!}\phi^4 \Bigr)_1
  \Bigl( -\frac{\lambda}{4!}\phi^4
  -\frac{(Z^2Z_{\lambda}-1)\lambda}{4!}\phi^4 \Bigr)_2 \Bigr] \nonumber \\
&+& T\Bigl[ \Bigl( -\frac{\lambda}{4!}\phi^4
    -\frac{(Z^2Z_{\lambda}-1)\lambda}{4!}\phi^4 \Bigr)_1
    \Bigl( \frac{(Z-1)}{2} \partial_{\nu}\phi\partial^{\nu}\phi\Bigr)_2 \Bigr]
\nonumber \\
&+& T\Bigl[ \Bigl( \frac{(Z-1)}{2} \partial_{\mu}\phi\partial^{\mu}\phi \Bigr)_1
\Bigl( -\frac{\lambda}{4!}\phi^4 -\frac{(Z^2Z_{\lambda}-1)\lambda}{4!}\phi^4
\Bigr)_2 \Bigr] \nonumber \\
&+& T\Bigl[ \Bigl( \frac{(Z-1)}{2} \partial_{\mu}\phi\partial^{\mu}\phi\Bigr)_1
  \Bigl( \frac{(Z-1)}{2} \partial_{\nu}\phi\partial^{\nu}\phi \Bigr)_2 \Bigr]
\nonumber \, .
\ea
We rewrite such expression by applying Wick theorem and then expressing the resulting
terms by means of the modified $T$-product using
(\ref{That1_phi}),~(\ref{That2_phi}). If we neglect terms with two contractions contributing to vacuum
diagrams, we get
\ba
T\Bigl[\Bigl( -{\cal L}' \Bigr)_1 \Bigl( -{\cal L}'\Bigr)_2\Bigr]
&=& \hat T\Bigl[ \Bigl( -\frac{\lambda}{4!}\phi^4
  -\frac{(Z^2Z_{\lambda}-1)\lambda}{4!}\phi^4 \Bigr)_1
  \Bigl( -\frac{\lambda}{4!}\phi^4
  -\frac{(Z^2Z_{\lambda}-1)\lambda}{4!}\phi^4 \Bigr)_2 \Bigr] \nonumber \\
&+& \hat T\Bigl[ \Bigl( -\frac{\lambda}{4!}\phi^4
    -\frac{(Z^2Z_{\lambda}-1)\lambda}{4!}\phi^4 \Bigr)_1
    \Bigl( \frac{(Z-1)}{2} \partial_{\nu}\phi\partial^{\nu}\phi\Bigr)_2 \Bigr]
\nonumber \\
&+& \hat T\Bigl[ \Bigl( \frac{(Z-1)}{2} \partial_{\mu}\phi\partial^{\mu}\phi
  \Bigr)_1
\Bigl( -\frac{\lambda}{4!}\phi^4 -\frac{(Z^2Z_{\lambda}-1)\lambda}{4!}\phi^4
\Bigr)_2 \Bigr] \nonumber \\
&+& \hat T\Bigl[ \Bigl( \frac{(Z-1)}{2}\partial_{\mu}\phi\partial^{\mu}\phi
  \Bigr)_1
  \Bigl( \frac{(Z-1)}{2} \partial_{\nu}\phi\partial^{\nu}\phi \Bigr)_2 \Bigr]
\nonumber \\
&+& N\left[\frac{(Z-1)^2}{4} (-i) \cdot 4 \cdot (\dot\phi_1)^2 \delta^{(4)}(x_1-x_2)
  \right] \nonumber \\
&=&  \hat T\Bigl[\Bigl( -{\cal L}' \Bigr)_1 \Bigl( -{\cal L}'\Bigr)_2\Bigr]
+ N\left[(Z-1)^2 (-i)(\dot\phi_1)^2 \delta^{(4)}(x_1-x_2) \right] \nonumber \;,
\ea
where the last, non-Lorentz-invariant term cancels exactly the non-invariant
term in $S^{(1)}$ of order $(Z-1)^2$, so that
\be
S^{(1)}+S^{(2)}=-i \int {d^4x_1\Bigl( -{\cal L}' \Bigr)_1}-\frac{1}{2}
\int {d^4x_1 d^4x_2 \hat T\Bigl[\Bigl( -{\cal L}' \Bigr)_1
    \Bigl( -{\cal L}'\Bigr)_2 \Bigr]}+O((Z-1)^3) \;.
\ee
The procedure can be repeated also for terms proportional to $(Z-1)^3\sim \lambda ^3$, which
requires considering $S^{(1)}$, $S^{(2)}$ and $S^{(3)}$, the latter being
given by
\begin{equation}
\begin{split}
  S^{(3)}&=\frac{i}{6} \int {d^4x_1 d^4x_2 d^4x_3 T\Bigl[ {\cal H}'(x_1)
      {\cal H}'(x_2) {\cal H}'(x_3) \Bigr]}\\
  &=\frac{i}{6} \int {d^4x_1 d^4x_2 d^4x_3 T\Bigl[\Bigl( -{\cal L}'
      + \frac{(Z-1)^2}{2Z}\dot{\phi}^2\Bigr)_{1} \Bigl( -{\cal L}'
      + \frac{(Z-1)^2}{2Z}\dot{\phi}^2\Bigr)_{2} \Bigl( -{\cal L}'
      + \frac{(Z-1)^2}{2Z}\dot{\phi}^2\Bigr)_{3}\Bigr]} \;.
\end{split}
\end{equation}
A straightforward, but tedious calculation, based on the application of
Wick theorem and then of equations (\ref{That1_phi}) and~(\ref{That2_phi}), leads to
\be
S^{(2)}=-\frac{1}{2}\int d^4x_1 d^4x_2 \hat T\Bigl[\Bigl( -{\cal L}'
      + \frac{(Z-1)^2}{2Z}\dot{\phi}^2\Bigr)_{1} \Bigl( -{\cal L}'
      + \frac{(Z-1)^2}{2Z}\dot{\phi}^2\Bigr)_{2} \Bigr ] + i \frac{(Z-1)^2}{2Z^2} \int d^4x_1
      \dot{\phi}^2_1
\ee
and also to
\ba
&& \frac{i}{6} \int d^4x_1 d^4x_2 d^4x_3 T\Bigl[\Bigl( -{\cal L}' \Bigr )_1
\Bigl( -{\cal L}' \Bigr )_2 \Bigl( -{\cal L}' \Bigr )_3 \Bigr ] =
\frac{i}{6} \int d^4x_1 d^4x_2 d^4x_3 \hat T\Bigl[\Bigl( -{\cal L}' \Bigr )_1
\Bigl( -{\cal L}' \Bigr )_2 \Bigl( -{\cal L}' \Bigr )_3 \Bigr ] \nonumber \\
&& + \frac{(Z-1)^2}{2} \int d^4x_1 d^4x_2 \hat T \Bigl[\Bigl( -{\cal L}' \Bigr )_1
 \dot{\phi}^2_2  \Bigr ] +i \frac{(Z-1)^3}{2} \int d^4x_1
      \dot{\phi}^2_1 \, .
 \ea
 In getting this last expression, we neglected double contractions insisting on
 the same couple of variables, which contribute to disconnected graphs,
 containing a vacuum diagram, and also a triple contraction, which produces a
 vacuum diagram.

Then, using the fact that both $Z-1$ and ${\cal L}'$ are $O(\lambda)$, we get
\ba
S^{(1)}+S^{(2)}+S^{(3)}&=&-i \int {d^4x_1\Bigl( -{\cal L}' \Bigr)_1}
-\frac{1}{2} \int {d^4x_1 d^4x_2 \hat T\Bigl[\Bigl( -{\cal L}' \Bigr)_1
    \Bigl( -{\cal L}'\Bigr)_2 \Bigr]} \nonumber \\
&+& \frac{i}{6} \int {d^4x_1 d^4x_2 d^4x_3\hat T\Bigl[\Bigl( -{\cal L}' \Bigr)_1
    \Bigl( -{\cal L}'\Bigr)_2 \Bigl(-{\cal L}' \Bigr)_3\Bigr]}+O(\lambda ^4) \;.
    \label{S3formula}
\ea
We stress that the equivalence~(\ref{S3formula}) holds if one neglects diagrams
containing a vacuum to vacuum process. This subtlety, even if not relevant for
practical applications, was apparently overlooked by~\cite{Nambu,Nishijima}.

\section{Conclusions}
\label{conclusions}
\setcounter{equation}{0}

In this short note we have given evidence that perturbative Dyson series for
the $S$-matrix enjoys relativistic invariance even in the case in which the
interaction depends on derivatives of the fields. This problem is usually
overlooked, since people almost always resort to Feynman diagrams and rules
which manifestly keep Lorentz invariance and which come naturally if field
theories are quantised by means of a functional generator. However the
equivalence between Feynman approach and the more traditional Dyson series is
not {\it a priori} evident if the interaction Lagrangian contains derivatives
of the fields. We have tackled this problem in the case of scalar electrodynamics
and renormalised $\phi ^4$ theory, giving simple perturbative arguments based on Wick theorem
that Dyson series for Hamiltonian is Lorentz invariant and - after discarding vacuum diagrams - coincides with perturbative series coming from applying Feynman rules to the Lagrangian. 
More in general, we are confident that by similar techniques the same coincidence can be
proven for all other renormalised quantum field theories, in particular for
QED.

\appendix

\section{Proof of (\ref{heis-int}) and~(\ref {heis-int-scal}) }
\label{appA}
\setcounter{equation}{0}

We give a proof of relations (\ref{heis-int}) and~(\ref {heis-int-scal}).
We start from the definition of a field in interaction representation:
\be
\phi _I=U \phi U^{-1} \, , \quad U=e^{iH_0^{(s)}t}e^{-iH^{(s)}t}\; ,
\ee
with $\phi $ in Heisenberg representation.
Then, we have
\be
\partial _0 U=-i e^{iH_0^{(s)}t}H'^{(s)}e^{-iH^{(s)}t}=-i U H'\;,
\ee
with
\be
H'=e^{iH^{(s)}t}H'^{(s)}e^{-iH^{(s)}t}
\ee
the interaction Hamiltonian in Heisenberg representation. As an immediate consequence we have
\be
\quad \partial _0 U^{-1}=i H' U^{-1}\;,
\ee
In addition, one has
\be
\partial _0 \phi =-i [\phi, H]\;,
\ee
for fields in Heisenberg representation.

Putting everything together we have
\be
\partial _0 \phi _I =-i U [\phi ,H_0] U^{-1} \, ,
\ee
with $H_0$ the 'free' Hamiltonian in Heisenberg representation.
For scalar electrodynamics it equals
\be
H_0=\int d^3 x {\cal H}_0 \, , \quad {\cal H}_0=
   \pi^\dag\pi +{\bm{\nabla}}\phi^\dag {\bm{\nabla}}\phi
   + m^2\phi^\dag\phi  + \frac{1}{4}F_{\mu\nu}F^{\mu\nu} \, .
\ee
For renormalised $\phi ^4$ theory it is
\be
   {\cal H}_{0}=\frac{\pi^2}{2}+\frac{1}{2}(\bm{\nabla}\phi)^2 \, .
\ee
Since in both cases we have the fundamental equal time commutation relation
(in the case of renormalised fields canonical commutation relations have to
be imposed on renormalised fields and momenta)
\be
[\phi (x), \pi (y) ] =i \delta ^3 (\vec{x}-\vec{y}) \, ,
\ee
we have
\be
\partial _0 \phi _I =-i U [\phi ,H_0] U^{-1}=U\pi ^{\dagger} U^{-1} \, ,
\ee
which gives immediately relations~(\ref{heis-int}) and~(\ref {heis-int-scal}), since for renormalised $\phi ^4$ theory $\pi ^{\dagger}=\pi$.

\end{document}